# Microwave Photonic Imaging Radar with a Millimeter-level Resolution

Cong Ma, Yue Yang, Ce Liu, Beichen Fan, Xingwei Ye, Yamei Zhang, Xiangchuan Wang, and Shilong Pan

*Abstract*—Microwave photonic radars enable fast or even real-time high-resolution imaging thanks to its broad bandwidth. Nevertheless, the frequency range of the radars usually overlaps with other existed radio-frequency (RF) applications, and only a centimeter-level imaging resolution has been reported, making them insufficient for civilian applications. Here, we propose a microwave photonic imaging radar with a millimeter-level resolution by introducing a frequency-stepped chirp signal based on an optical frequency shifting loop. As compared with the conventional linear-frequency modulated (LFM) signal, the frequency-stepped chirp signal can bring the system excellent capability of anti-interference. In an experiment, a frequency-stepped chirp signal with a total bandwidth of 18.2 GHz (16.9 to 35.1 GHz) is generated. Postprocessing the radar echo, radar imaging with a two-dimensional imaging resolution of ~8.5 mm×~8.3 mm is achieved. An auto-regressive algorithm is used to reconstruct the disturbed signal when a frequency interference exists, and the high-resolution imaging is sustained.

## I. INTRODUCTION

Radars have the capability of providing weather independent and day-and-night detection, which play more and more important roles in civilian applications [1]. To realize accurate target feature extraction, ultra-high-resolution imaging is extremely desirable in applications like foreign objects debris (FOD) detection on runways, security check, autonomous drive and so on [2-4]. In imaging radar systems, broadband waveforms are required to realize ultra-high range resolution [5, 6]. However, generation of the broadband signals in the electronic domain requires multi-stage frequency multiplication and up-conversions, which would deteriorate the signal quality and significantly increase the system complexity [7]. Even if the broadband waveform can be produced, it is challenging for electronic approaches to process it with small latency. Microwave photonics is a feasible solution to the problem mentioned above for its distinct features of wide bandwidth, low transmission loss and excellent electromagnetic interference immunity [8-16]. Previously, a digital radar with a photonic transceiver was demonstrated using a passive mode-locked laser (MLL) with ultra-low timing jitter [17], which showed the feasibility of microwave photonic radars for practical applications. However, the bandwidth was limited to several tens of MHz and the range resolution was in the order of meters due to the low repetition rate of the passive MLL. To improve the bandwidth, a variety of microwave photonic radar architectures were proposed. By applying wavelength-to-time mapping, a chirp signal with a bandwidth up to 30 GHz was obtained [18-21], but the time-bandwidth product of the generated chirp signal is rather small, which is not suitable for radar applications with long-distance detection. In addition, de-chirping of the broadband signal with a very small time duration may output a signal with a frequency of several GHz, leading to a huge amount of data after the analog-to-digital converter (ADC). A logic-operation-based photonic digital-to-analog converter, which was applied to a W-band microwave photonic imaging radar, could generate a chirp signal with a bandwidth of 8 GHz, leading to a range resolution of 1.9 cm [22]. However, the bandwidth was limited by the bit rate of the digital drive signal, making it difficult to further improve the range resolution. Broadband chirp signals generated via microwave photonic frequency multiplication were also adopted in microwave photonic imaging radars [23-25], where a bandwidth of 12 GHz and a range resolution of 1.3 cm were achieved [26]. Further improving the range resolution to the millimeter-level requires the generation and processing of a chirp signal with an instantaneous bandwidth of larger than 15 GHz. However, similar to the electronic methods, many unwanted harmonics that cannot be eliminated by filters will be generated, and the signal-to-noise ratio of the signal will be degraded.

The aforementioned microwave photonic imaging radars with high-resolution mainly adopt a chirp signal, also known as linear-frequency modulation (LFM) signal, which is one of the most widely used waveforms in wideband radars due to its excellent pulse compression capability [27, 28]. By de-chirp processing the echos, a signal with a frequency that is much smaller than the original chirp signal is achieved, so a low-speed ADC is sufficient to sample the signal in the receiver. Fast or even real-time imaging is therefore enabled because the amount of data is small [29, 30]. However, with the increase of the radar's operating bandwidth, the frequency ranges of the radars may overlap with other applications, such as communication devices and automotive radars. A frequency-stepped

This work was supported in part by the National Key R&D Program of China (2018YFB2201803), National Natural Science Foundation of China (61901215), and Jiangsu Provincial "333" Project (BRA2018042).

The authors are with the Key Laboratory of Radar Imaging and Microwave Photonics (Nanjing Univ. Aeronaut. Astronaut.), Ministry of Education, Nanjing University of Aeronautics and Astronautics, Nanjing 210016, China (e-mail: pans@nuaa.edu.cn).

signal with intra-pulse linear frequency modulation, also known as frequency-stepped chirp signal, has both advantages of the chirp signal and the frequency-stepped signal [28], which may be employed to achieve high-resolution detection under interference scenario. Although researchers in both the electrical and optical areas have tried the frequency-stepped signal in radar systems [31-33], only few segments of chirp signals were generated and exploited, which presented with a limited resolution.

In this paper, we propose and experimentally demonstrate a microwave photonic imaging radar with a millimeter-level two-dimensional resolution, in which a frequency-stepped chirp signal is generated and processed in the optical domain. By utilizing an optical frequency shifting loop (OFSL) based on carrier-suppressed single sideband (CS-SSB) modulation [34], we successfully generate and transmit nine segments of sub-pulse with an individual bandwidth of 2.2 GHz and a center frequency difference of 2 GHz between two adjacent subpulses. Much fewer unwanted harmonics are produced in this work than the conventional microwave photonic frequency multiplication method. In the receiver, postprocessing is applied to extract high-resolution target features. Because optical mixers can process signals with large bandwidths, we adopt a bandwidth synthesis algorithm based on the de-chirp receiving mode to achieve fast data fusion [35, 36]. A signal equivalent to the result de-chirped from one chirp signal with a bandwidth of 18.2 GHz is obtained by the bandwidth synthesis algorithm, and radar imaging with a resolution of ~8.5 mm×~8.3 mm is achieved. Considering the situation of existence of interference, we can remove the disturbed signal and reconstruct the missing data by using the auto-regressive (AR) algorithm thanks to the abundant priori information introduced by the large operating bandwidth of the system [37, 38], bring the system excellent capability of anti-interference.

## II. PRINCIPLE

Fig. 1 shows the schematic diagram of the proposed microwave photonic radar system. A continuous-wave (CW) light with a frequency of $f_0$ from a laser diode (LD) is divided into two branches by an optical coupler (OC, OC1). In the upper branch, a pulsed light with a period of $T_{pr}$ and a pulse width of $T_{pw}$ is generated through an optical switch driven by an electrical pulse train. The pulsed light, serving as the optical seed source, is injected into an OFSL, which mainly comprises a dual-parallel Mach-Zehnder modulator (DPMZM), an optical amplifier (OA), and an optical bandpass filter (OBPF). The DPMZM, which is an integrated device consisting of two sub-MZMs (MZM-a and MZM-b) and a parent-MZM (MZM-c), is driven by two radio-frequency (RF) signals with a fixed phase difference of 90°, as shown in Fig. 1(b). Considering the small-signal modulation condition, CS-SSB modulation (+1st-order sideband) can be achieved by biasing MZM-a and MZM-b at the null transmission point, and MZM-c at the quadrature transmission point. The OA is used to compensate for the loop loss, and the OBPF is used to control the bandwidth of the OFSL and to reduce the spontaneous emission from the OA. Assuming that the frequency of the RF signal is $\Delta f$, the optical signal in the loop will frequency shift from the previous one by $\Delta f$ in each cycle. As a result, the optical pulsed signal coupled out from OC3 can be expressed as

$$E_{\text{OFSL}} = E_{\text{PL}} \sum_{n=0}^{N_{\max}-1} \text{rect}(\frac{t-nT_{\text{L}}}{T_{\text{pw}}}) \cdot \exp\{j2\pi[f_0 + (n+1)\Delta f](t-nT_{\text{L}}) + \varphi_n\} \quad (1)$$

where $E_{\text{PL}}$ is the amplitude of the pulsed light, $N_{\max}$ is the maximum number of frequency-shifted pulses in a period, $T_{\text{L}}$ is the circulation time of the loop, $\varphi_n$ is the initial phase of the $n$th optical pulsed signal, and rect($x$) is the rectangular function, which is given by rect($x$)={1, $|x| \leq 0.5$; 0, $|x| > 0.5$}. The maximum number of optical pulsed signals in a period is mainly limited by the bandwidth $B_{\text{OBPF}}$ of the OBPF, and $N_{\max}=[B_{\text{OBPF}}/\Delta f]+1$, where [$x$] denotes the largest integer no more than $x$.

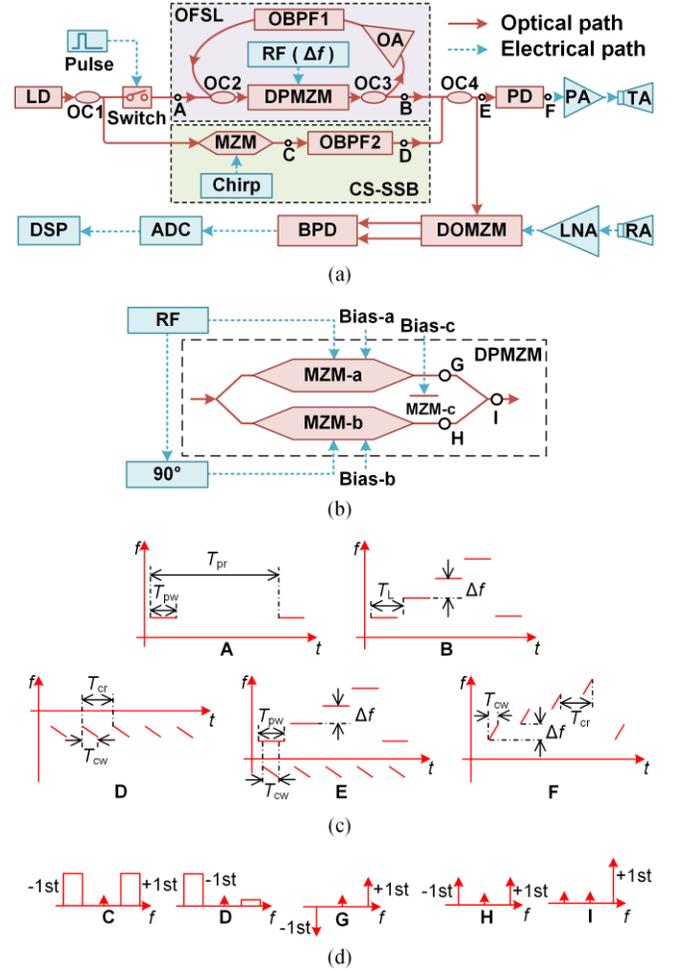

Fig. 1. (a) Schematic diagram of the proposed microwave photonic radar system. (b) Configuration of the DPMZM. (c) Frequency-time diagrams and (d) optical spectra at several key points. LD, laser diode; OC, optical coupler; MZM, Mach-Zehnder modulator; DPMZM, dual-parallel MZM; DOMZM, dual-output MZM; OA, optical amplifier; OBPF, optical bandpass filter; RF, radio-frequency; CS-SSB, carrier-suppressed single-sideband; PD, photodetector; BPD, balanced PD; PA, power amplifier; TA, transmit antenna; RA, receive antenna; LNA, low noise amplifier; ADC, analog-to-digital converter; DSP, digital signal processor.

In the lower branch, the CW light is modulated by a narrowband electrical chirp signal through CS-SSB modulation. Here, the CS-SSB modulation is implemented by cascading an MZM and an OBPF, where the MZM is biased at the minimum transmission point to perform carrier-suppressed double sideband modulation, and the −1$^{st}$-order sideband is filtered out by the OBPF. Assuming the center frequency and the chirp rate of the chirp signal are $f_c$ and $k$, the output optical signal in the lower branch is written as

$$E_{SSB} = E_{CW} \sum_n \text{rect}(\frac{t - nT_{cr}}{T_{cw}}) \cdot \exp\{j2\pi[f_0 - f_c - 1/2k(t - nT_{cr})](t - nT_{cr})\} \quad (2)$$

where $E_{CW}$ is the amplitude of the optical signal, $T_{cr}$ and $T_{cw}$ are the periods and the pulse widths of the chirp signals.

The two branches are combined again via OC4 and further divided into two parts. One portion is sent to a photodetector (PD) to generate a frequency-stepped chirp signal, which can be given by

$$I(t) \propto \sum_{n=0}^{N_{max}-1} \text{rect}(\frac{t - nT_{cr}}{T_{cw}}) \cdot \cos\{2\pi[f_{cn} + 1/2k(t - nT_{cr})](t - nT_{cr})\} \quad (3)$$

where $f_{cn}=f_c+(n+1)\Delta f$. Afterward, the frequency-stepped chirp signal is amplified via an electrical power amplifier (PA) and then sent to the free space via a transmit antenna.

It should be noted that each optical pulsed signal generated by the OFSL should temporally coincide with that generated by the CS-SSB module in the lower branch to generate a suitable frequency-stepped chirp signal, as shown in the frequency-time diagram E of Fig. 1(c), so the conditions of $T_{pr}=MT_{cr}=MT_L$ and $T_{pw} \geq T_{cw}$ should be satisfied, where $M$ is an integer. In addition, it is necessary to ensure that there are no residual optical pulse signals in the OFSL before the optical pulse signal is injected into the loop, i.e., the trailing edge of the previous optical pulse signal has left the OFSL before the leading edge of the optical pulse signal is injected into the loop, so $M \geq N_{max}$ and $T_L \geq T_{pw}$.

The other port of OC4 is connected to a dual-output MZM (DOMZM), which is biased at the quadrature transmission point. The radar echo gathered by the receive antenna is augmented via a low-noise amplifier (LNA) and is exploited to drive the DOMZM. Assuming that the temporal delay between the reference signal and the echo signal is $\tau$. After the DOMZM, a frequency pair at $f_0+(n+1)\Delta f$ and $f_0+(n+1)\Delta f-k\tau$, and another frequency pair at $f_0-f_c-k(t-T_{cr})$ and $f_0-f_c-k(t-T_{cr}-\tau)$ are generated. The output optical signals of the DOMZM are detected by a low-frequency balanced photodetector (BPD) with a narrow bandwidth, so that the undesired high-frequency signals could not be generated. The de-chirped signal generated at the BPD can be expressed as

$$s(t) \propto \sum_{n=0}^{N_{max}-1} \text{rect}(\frac{t - nT_{cr} - \tau}{T_{cw} - \tau}) \cdot \cos\{2\pi[k\tau(t - nT_{cr}) - 1/2k\tau^2 + f_{cn}\tau]\} \quad (4)$$

For convenience, the $n$th de-chirped sub-signal in (4) is written as

$$s_n(t) \propto \cos\{2\pi[k\tau(t - nT_{cr}) - 1/2k\tau^2 + f_{cn}\tau]\} \\ = \cos\{2\pi[k\tau(t - nT_{cr}) - 1/2k\tau^2 + f_{c(n-1)}\tau + \Delta f\tau]\} \quad (5)$$

Time-shifting $s_n(t)$, we have

$$s_n(t + T_{cr} - \Delta f/k) \\ \propto \cos\{2\pi\{k\tau[t - (n-1)T_{cr}] - 1/2k\tau^2 + f_{c(n-1)}\tau\}\} \quad (6)$$

As can be seen, the expression of $s_n(t+T_{cr}-\Delta f/k)$ is exactly the same as that of $s_{n-1}(t)$, which means that the de-chirped sub-signals are coherent. After a suitable time-shift, the de-chirped sub-signals can be combined into one signal in the time domain. The synthetic de-chirped signal is equivalent to a signal de-chirped from a chirp signal with a large bandwidth, i.e., bandwidth synthesis is achieved.

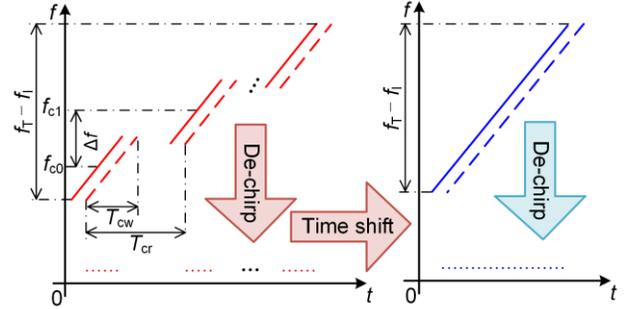

Fig. 2. The principle of bandwidth synthesis. Solid line, transmitted signals; dash line, received signals; dot line, de-chirped signals.

The principle of bandwidth synthesis can be illustrated in Fig. 2. First, we de-chirp $N$ subpulses with stepped center frequencies to obtain $N$ low-frequency de-chirped sub-signals. Then, the $N$ low-frequency sub-signals are combined together to obtain a signal equivalent to that de-chirped from a chirp signal with a bandwidth of $f_T-f_I$, where $f_T$ and $f_I$ are the terminal frequency and the initial frequency of the $N$ subpulses. It should be noted that the case of $\Delta f < B_{chirp}$ should be satisfied, where $B_{chirp}$ is the individual bandwidth of the subpulse. Otherwise, it would generate discontinuous phases at the joint points of the synthetic signal because the sub-signal duration after de-chirping is less than $T_{cw}$.

To achieve an ultra-high range resolution, a large operating bandwidth is desired. However, due to the scarcity of spectrum resources, the frequency ranges of wideband radars may overlap with other existed RF applications, causing the radar to malfunction if interference exists. With a known frequency band of the interference signal, the frequency interval $\Delta f$ between the subpulses can be adjusted to avoid the interference band, or the data in the corresponding subpulse can be discarded. These operations will lead to gaps in the synthetic signal, resulting in phase discontinuities and creating false peaks in the frequency domain. In order to conduct high-resolution imaging, this missing data must be reconstructed. Thanks to the large operation bandwidth of the proposed radar, much more priori information can be obtained. The signals on both sides of the signal gap are de-chirped from the subpulses in the lower and upper bands, and represented using the AR model [37, 38]. The parameters of the model are estimated by using the data of the upper and lower de-chirped signals, and the model is adjusted to match the signal optimally. Then, the signal gap is accurately filled through the adjusted model, as shown in Fig.3. Therefore, the proposed radar can work normally under interference.

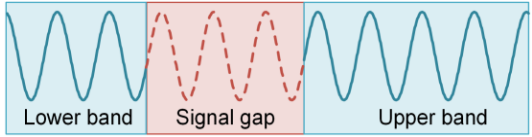

Fig. 3. Illustration of the signal gap filling after de-chirping.

After analog-to-digital conversion, the low-frequency de-chirped signals can be processed quickly in the digital signal processor (DSP). It is well known that the range resolution $R$ of radar is determined by bandwidth $B$ of the signal via $R=c/2B$, where $c$ is the wave propagation velocity [5]. Therefore, the proposed microwave photonic radar can achieve an ultra-high range resolution because large equivalent bandwidth can be obtained by combining the sub-signals de-chirped from a number of narrowband subpulses. If an inverse synthetic aperture radar (ISAR) imaging algorithm is used, fast or even real-time radar imaging with an ultra-high resolution can be implemented [6].

### III. EXPERIMENTS AND RESULTS

An experiment is carried out based on the setup in Fig.1. A light wave at 1550.1 nm ($f_0$=193.399 THz) with a power of 16 dBm from a laser source (TeraXion PureSpectrum-NLL) is equally split into two branches by OC1. The upper branch is sent to a 200-MHz acousto-optic modulator (AOM), which acts as a switch and is driven by an electrical pulse with a period of 71.96 μs and a pulse width of 5 μs. The output spectrum of the AOM is shown as the blue curve in Fig. 4(a). When the electrical pulse is at a high level, the AOM outputs an optical signal with a frequency shift of 200 MHz. On the contrary, when the electric pulse is at a low level, the AOM has no optical signal output. Thus a pulsed light with a period of 71.96 μs and a pulse width of 5 μs is generated. The pulsed light is injected into the OFSL, which consists of a DPMZM (Fujitsu FTM7961), an erbium-doped fiber amplifier (EDFA, Amonics AEDFA-PA-35B), an OBPF (Yenista XTM-50), two OCs and a 1-km fiber to achieve a circulation time of 5.14 μs. A 2-GHz RF signal with a power of 13 dBm generated by a signal generator (Agilent E8257D) passes through a 90° hybrid before it is applied to the DPMZM working at the CS-SSB modulation mode. When the OFSL is open, the output optical spectrum of the DPMZM is shown as the red curve in Fig. 4(a). As can be seen, the frequency of the output is increased by 2 GHz as compared with the input. The bandwidth of the OBPF is set to 16 GHz, as shown as the blue curve in Fig. 4(b). After carefully adjusting the pump power of the EDFA, the loop loss is compensated. The output optical spectrum of the OFSL is shown by the red curve in Fig. 4(b), in which 9 optical comb lines with a power fluctuation of less than 1 dB are observed.

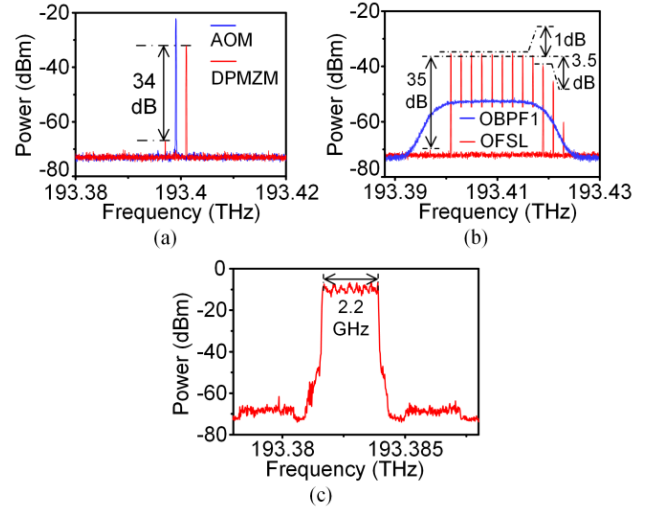

Fig. 4. (a) The output optical spectra of the AOM and the DPMZM. (b) The response of OBPF1 and the output optical spectrum of the OFSL. (c) The output optical spectrum of the CS-SSB module.

The lower branch of OC1 is connected to an MZM, which is biased at the null transmission point. A 2.2-GHz-bandwidth chirp signal (14.7 to 16.9 GHz) with a period of 5.14 μs and a pulse width of 3.3 μs generated by an arbitrary waveform generator (AWG, Keysight 8195A) are introduced to the RF port of the MZM. The chirp signal from the AWG output is captured by a 32-GHz real-time oscilloscope (Keysight 93304A) with a maximum sampling rate of 80 GSa/s, and short-time Fourier transform (STFT) analysis is performed to the recorded data, leading to a frequency-time diagram shown in Fig. 5(a1). The corresponding spectrum is measured by an electrical spectrum analyzer (Keysight N9010A, bandwidth 44 GHz) with a resolution bandwidth (RBW) of 3 MHz, with the result shown in Fig. 5(b1). The MZM and a following OBPF form a CS-SSB module, and the output optical spectrum is shown in Fig. 4(c). The signal from the CS-SSB module and the optical signal from the OFSL are amplified, combined and further split into two branches. One branch is sent to a 40-GHz PD with a responsivity of 0.65 A/W to generate the frequency-stepped chirp signal, which is amplified by a PA before being sent to a horn antenna for radiation. The frequency-time diagram and the electrical spectrum of the generated frequency-stepped chirp signal are shown in Figs. 5(a2) and 5(b2). Nine subpulses covering a frequency range from 16.9 GHz to 35.1 GHz with a center frequency interval of 2 GHz can be seen in Fig. 5(a2). The ninth subpulse is severely distorted due to the limited bandwidth of the oscilloscope. It is worth noting that several weak harmonic signals are generated due to the inad-equate suppression of the optical carrier by the DPMZM in the OFSL. The performance deterioration of the radar system caused by these harmonic signals can be ignored in the experiment because its power is very weak. By employing a DPMZM with a higher extinction ratio, the power of the harmonic components can be further decreased.

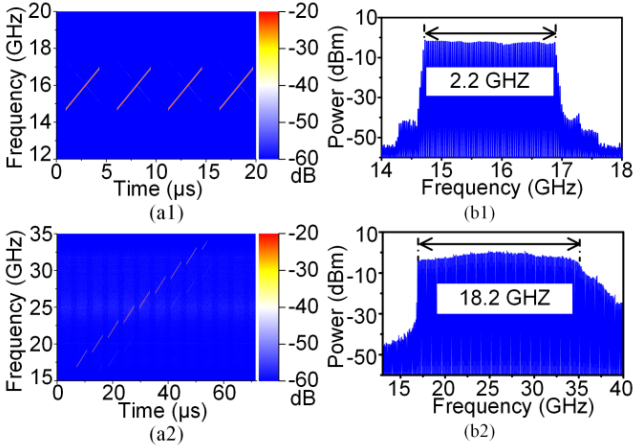

Fig. 5. (a) The frequency-time diagrams and (b) the electrical spectra of (1) the input chirp signal and (2) the output frequency-stepped chirp signal.

The other branch of OC4 is connected to the optical input port of a DOMZM (Eospace AX-1x2-0MVS). The echoes collected via the receive antenna are augmented by an LNA and then fed to the RF port of the DOMZM, which is biased at the quadrature transmission point. Then, the optical signal modulated by the electric echo signal beats at a BPD (Thorlabs PDB450C-AC) embedded with LNAs to generate a de-chirped signal. The de-chirped signal is further converted into a digital signal by the real-time oscilloscope. Since the frequency is low, a sampling rate of 100 MSa/s is exploited. A suitable time shift is introduced to the component de-chirped from the individual subpulse via a digital processing algorithm to achieve bandwidth synthesis.

To investigate the range resolution of the proposed radar system, a cable is used to connect the PD and the DOMZM via an LNA to simulate the detection of a single target while isolating the impact of the external environment. According to (6) the de-chirped sub-signals are time-shifted and combined to obtain synthetic signals. By applying the fast Fourier transform (FFT), the spectrum of the synthesized signal after de-chirping is obtained. Fig. 6(a) shows the obtained spectra when the number $N$ of de-chirped sub-signals used in the synthesis is 3, 6, and 9 (i.e., the corresponding equivalent bandwidth is 6.2 GHz, 12.2 GHz, and 18.2 GHz), respectively. As can be seen, the more de-chirped sub-signals are used in the synthesis, the narrower the main lobe is, i.e., the higher the range resolution is. The range resolution of the radar system can be obtained by measuring the 3.92-dB bandwidth of the main lobe in the spectrum of the de-chirped signal. For an ideal case, the 3.92-dB bandwidth could lead to the best range resolution of $R=c/2B$ [5]. Fig. 6(b) shows the comparison of theoretical and experimental results of range resolution when different numbers of de-chirped sub-signals are used in the synthesis. As illustrated, the two lines match very well. When the number of signals used in the synthesis is 9, the 3.92 dB bandwidth of the main lobe is 37.4 kHz, and the corresponding range resolution is calculated to be 8.4 mm, which is an error of 0.2 mm from the theoretical resolution (8.2 mm).

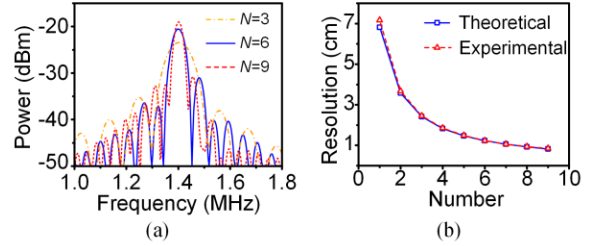

Fig. 6. (a) The spectra of the synthesized de-chirped signals when different numbers of de-chirped sub-signals are used in the synthesis. (b) The theoretical and experimental resolutions as a function of the number of de-chirped sub-signals.

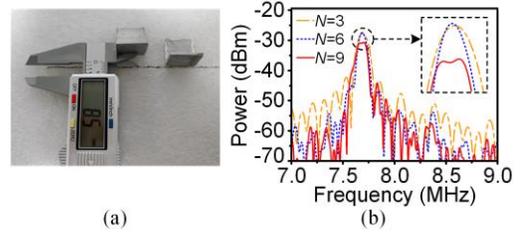

Fig. 7. (a) The picture of the targets. (b) The measured spectra when different numbers of de-chirped sub-signals are used in the synthesis.

Then, two metallic plane targets (2 cm×2 cm) spaced 8.5 mm apart, as shown in Fig. 7(a), are detected by the proposed radar to calibrate the range resolution of the radar. Fig. 7(b) shows the obtained spectra when the number of de-chirped sub-signals used in the bandwidth synthesis is 3, 6, and 9 (i.e., the corresponding equivalent bandwidth is 6.2 GHz, 12.2 GHz, and 18.2 GHz). Two peaks located at 7.665 MHz and 7.703 MHz can be observed, whereas only one peak can be observed in other curves. The frequency difference between the two peaks is 38 kHz, corresponding to a measured distance of 8.52 mm, which shows that the two targets can be distinguished with an error of 0.02 mm, validating again the millimeter-level range resolution of the proposed radar.

To further investigate the radar detection performance, ISAR imaging experiments are conducted. The target is a slanted letter "V" with a side length of 9.8 cm and an angle of 53° composed of aluminum particles, which are placed on a rotator with a rotating speed of 360°/s. The distance between the rotator and the antennas is about 1.5 m. The picture of the target is shown in Fig. 8(a). The frequencies of the de-chirped signals are located at around 7.5 MHz and the sampling time is set to 0.11 s. Under this setting, the viewing angle for every ISAR picture is 39.6°, and the cross-range resolution is ~8.3 mm [6]. Thus, a 2-D imaging resolution of ~8.5 mm×~8.3 mm can be obtained when $N$=9. Figs. 8(b) to 8(d) show the ISAR images constructed when the number $N$ of the de-chirped sub-signals used in the synthesis is 3, 6, and 9 (i.e., the corresponding equivalent bandwidth is 6.2 GHz, 12.2 GHz, and 18.2 GHz). As can be seen, the resolution is improved significantly as *N* increases.

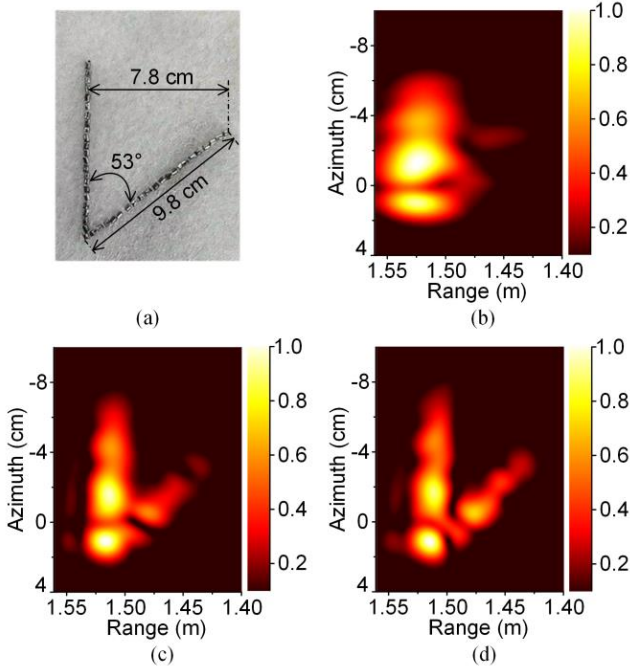

Fig. 8. (a) The picture of the target and results of the ISAR demonstration when (b) N=3, (c) N=6, (d) N=9.

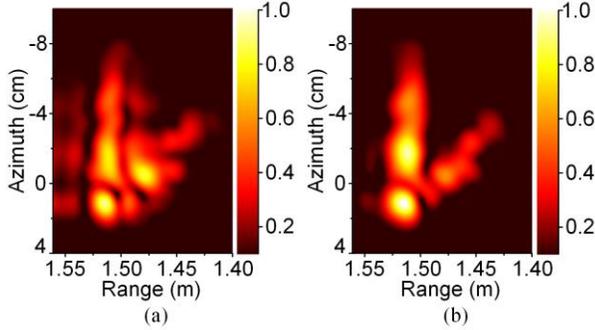

Fig. 9. The ISAR image when the radar is interfered. (a) Before and (b) after using the AR algorithm.

In addition, we test the situation assuming the radar is disturbed by an interference signal in the range of 22.9 to 25.1 GHz. To remove the interference, we set the amplitude of the fourth de-chirped sub-signal to zero. If the signal gap is not filled with the reconstructed data, it will create false peaks in the frequency domain, which in turn produce a very blurred ISAR image, as shown in Fig. 9(a). It should be noted that it is difficult to reconstruct the missing data when there is little prior information. Fortunately, the proposed system has a large bandwidth and can get enough prior information to reconstruct the missing data. The missing data can be reconstructed by the AR algorithm and filled into the signal gap. ISAR imaging can be applied to the gap-filled signals to obtain an ultra-high-resolution image, as shown in Fig. 9(b), which is almost the same as the previous image without interference. This shows that the proposed radar exhibits good anti-interference ability.

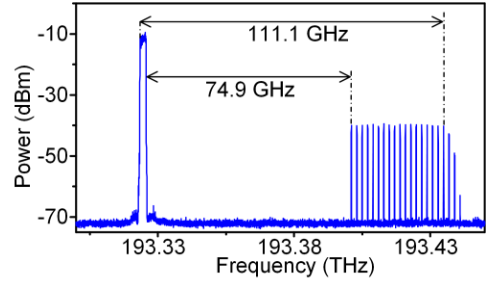

Fig. 10. The output optical spectra of OC4.

It is worth noting that the highest frequency of the frequency-stepped chirp signal is mainly limited by the PD bandwidth. High-frequency signals, such as W-band (75 to 110 GHz), can be generated by using a broadband PD. The total bandwidth is limited by the OBPF in the OFSL, so a larger bandwidth can be achieved by increasing the bandwidth of the OBPF in the OFSL. The two lights separated by OC1 can be replaced by two phase-locked lights with a frequency interval of 58 GHz, and the bandwidth of the OBPF in the OFSL can be increased to 34 GHz. After carefully adjusting the system, an optical spectrum as shown in Fig. 10 can be achieved at the output port of OC4. As can be seen, the system has the potential to generate a frequency-stepped chirp signal covering the entire W-band. After de-chirping and data fusion processing, a range resolution of 4.3 mm could be envisaged.

## IV. CONCLUSIONS

In conclusion, a millimeter-level resolution imaging radar has been proposed and demonstrated with a frequency-stepped chirp signal generated by an optical frequency shifting loop. An 18.2 GHz bandwidth and an 8.5 mm range resolution are achieved whereas the use of the large-bandwidth electronic devices to generate and process broadband chirp signals is avoided. Radar imaging with a two-dimensional (range and cross-range) imaging resolution of ~8.5 mm × ~8.3 mm is implemented, and the anti-interference ability of the system is verified. In addition, the system has the potential to generate W-band signals, which may enable radar imaging with a higher resolution.